\documentclass{aa}  
\usepackage[colorlinks=true,linkcolor=blue,citecolor=blue,breaklinks=true]{hyperref}
\usepackage{graphicx}
\usepackage{txfonts}
\usepackage{float}
\usepackage{rotating,tabularx}
\usepackage{pdflscape}

\usepackage{multirow}
\usepackage{afterpage}
\usepackage{bigstrut}
\usepackage[table,xcdraw]{xcolor}
\usepackage[flushleft]{threeparttable }
\usepackage{amsmath}
\usepackage{enumitem}
\usepackage{amsfonts,color}
\usepackage{amssymb,float}
\usepackage[utf8]{inputenc}
\usepackage{color}
\usepackage{etoolbox}
\usepackage{appendix}
\usepackage{url}
\usepackage{dirtytalk}
\usepackage{float}
\usepackage{fontawesome}

\usepackage{booktabs}
\usepackage[graphicx]{realboxes}
\usepackage{adjustbox}
\usepackage{multirow}
\usepackage{rotating,tabularx}
\usepackage{afterpage}
\usepackage{tablefootnote}

\usepackage{tikz}

\usepackage{bigstrut}

\definecolor{lime}{HTML}{A6CE39}
\DeclareRobustCommand{\orcidicon}{%
    \begin{tikzpicture}
    \draw[lime, fill=lime] (0,0) 
    circle [radius=0.16] 
    node[white] {{\fontfamily{qag}\selectfont \tiny ID}};
    \draw[white, fill=white] (-0.0625,0.095) 
    circle [radius=0.007];
    \end{tikzpicture}
    \hspace{-2mm}
}

\foreach \x in {A, ..., Z}{%
    \expandafter\xdef\csname orcid\x\endcsname{\noexpand\href{https://orcid.org/\csname orcidauthor\x\endcsname}{\noexpand\orcidicon}}
}
\newcommand{\orcid}[1]{\href{https://orcid.org/#1}{\textcolor[HTML]{A6CE39}{\orcidicon}}}
\usepackage{graphicx}
\usepackage{txfonts}

\newcommand{\gaia}{\textit{Gaia}}

\usepackage{arydshln}
\usepackage{dsfont}
\usepackage{makecell}
\begin{document} 
\title{A $1\%$ distance to the Large Magellanic Cloud measured by population-II pulsating stars using {\it Gaia} Data Release 3}
\titlerunning{A $1\%$ distance to the LMC from Population-II pulsators}

     \author{Bastian Lengen\inst{1}\orcid{0009-0007-8211-8262} \and Richard I. Anderson\inst{2,1}\orcid{0000-0001-8089-4419}  \and Mauricio Cruz Reyes \inst{1}\orcid{0000-0003-2443-173X}} 

   \institute{Institute of Physics, \'Ecole Polytechnique F\'ed\'erale de Lausanne (EPFL), Chemin Pegasi 51b, 1290 Versoix, Switzerland 
   \and
   Institut f\"ur Astrophysik und Geophysik, Georg-August-Universit\"at G\"ottingen, Friedrich-Hund-Platz 1, 37077 G\"ottingen, Germany
   \\
   \email{bastian.lengen@epfl.ch, richard.anderson@uni-goettingen.de}}

   \date{Submitted May 25, 2026} 

    \abstract{
    Population-II pulsating stars provide a route to extragalactic distances that is independent of the classical Cepheid distance scale and complementary to geometric and tip-of-the-red-giant-branch (TRGB) methods. We apply optical Wesenheit Leavitt laws for RR Lyrae and type II Cepheid stars calibrated with \gaia\ DR3 data and anchored by homogeneous globular-cluster distances from trigonometric parallaxes to variable stars in the Large Magellanic Cloud (LMC). We adopt RRab stars as the baseline tracer because they define the absolute zero point of the calibration, dominate the LMC sample, and provide robust classifications. The uncertainty budget propagates the full covariance of the calibration parameters, treating calibration uncertainties as correlated systematics rather than independent star-by-star errors. Using 12\,193 RRab stars after outlier rejection, we determine $\mu_{\mathrm{LMC}} = 18.423 \pm 0.002\,(\mathrm{stat}) \pm 0.020\,(\mathrm{syst})$\,mag. This combines the statistical uncertainty on the mean and the systematic uncertainty of the absolute calibration, which currently limits the total precision. Our result is lower than the detached-eclipsing-binary benchmark by $0.054$\,mag, corresponding to a $\sim1.7\sigma$ offset, and agrees with the TRGB distance obtained from the same globular-cluster scale to within $0.024$\,mag. RRc and T2Cep stars provide useful consistency checks, but the relative offsets RRc--RRab and T2Cep--RRab measured in the LMC differ from those calibrated in globular clusters. Geometric corrections between tracer barycenters and external reference positions are below $0.003$\,mag. Individual RRab distances map the three-dimensional structure of the LMC over a broad $10^\circ$-radius field. A planar model reproduces the dominant distance gradient and yields $i = 21.3 \pm 0.7^\circ$ and $\Theta = 145.2 \pm 2.2^\circ$, in agreement with previous determinations. 
    }

   
   \maketitle

\section{Introduction}\label{sec:introduction}
Pulsating stars provide a fundamental route to measure distances across the Universe through the period--luminosity relations (PLR), or Leavitt laws \citep[][LL]{leavitt1912}, that they follow. While classical Cepheids have long been the cornerstone of the distance ladder \citep[e.g.][]{riess2022}, population-II pulsators such as RR Lyrae (RRL) and type II Cepheid (T2Cep) stars offer an independent and complementary pathway to calibrate the extragalactic distance scale. Establishing a robust population-II distance scale is essential to provide cross checks on existing methods and to assess potential systematic uncertainties affecting the distance ladder.

Recent developments have led to a transition from a simple hierarchical distance ladder toward a more interconnected distance network \citep{H0DN2026}, in which multiple partially independent routes contribute to the overall calibration. In this framework, the availability of additional tracers and independent anchors to \gaia\ parallaxes provides critical redundancy and enables stringent tests of systematic effects. Population-II pulsators constitute a key component of this network, offering a route that is largely independent of classical Cepheids while being directly anchored to trigonometric parallaxes. In particular, RR Lyrae stars are powerful distance tracers of the Galactic halo, stellar streams, and nearby galaxies, making improvements in their absolute calibration highly relevant for near-field cosmology and the broader extragalactic distance scale.

RR Lyrae and type II Cepheid stars are well established standard candles that follow well-defined Leavitt laws, enabling distance determinations across the Local Group and beyond. These relations are observed in the infrared and in reddening-free magnitudes \citep[e.g.][]{bhardwaj2017, neeley2019, soszynski2019, braga2020, cruzreyes2025}. However, their calibration has historically been limited by the accuracy of available parallaxes and by the treatment of systematic effects such as metallicity dependencies and population differences. The advent of \gaia, in particular with its third data release \citep[GDR3]{gaiaDR32023, clementini2023, ripepi2023}, has significantly improved this situation by providing homogeneous astrometry and photometry for large samples of variable stars, thereby enabling a new generation of population-II calibrations \citep[e.g.][]{garofalo2022, lengen2026}.

In \citet[][hereafter L26]{lengen2026}, we presented a joint calibration of the Leavitt laws of fundamental-mode (RRab) and first-overtone (RRc) RR Lyrae stars, together with type II Cepheid (T2Cep) stars, based on GDR3 data. This approach relies on a global fit including all three types of pulsators across 93 globular clusters (GCs), with a homogeneous treatment of parallaxes and metallicities, enabling a self-consistent calibration anchored to these GCs. The resulting relations provide a precise and unified description of population-II pulsators, with distances calibrated at the percent level.

The Large Magellanic Cloud (LMC) plays a central role in the distance network as a primary anchor galaxy. Its distance is known with high precision from geometric methods, in particular from detached eclipsing binaries (DEB) measurements \citep{pietrzyski2019}, making the LMC an ideal benchmark for testing independent distance indicators. Complementary constraints are also provided by population-II tracers such as the tip of the red giant branch (TRGB). Using a calibration anchored to the same globular cluster sample and \gaia-based geometric distance scale as L26, \citet{cruzreyes2026} recently derived an independent TRGB distance to the LMC. The fact that both approaches share the same geometric anchoring, while relying on fundamentally different observables, provides a particularly valuable cross-check of the population-II distance scale: pulsating stars act as individual standard candles, whereas the TRGB is a statistical feature of the color--magnitude diagram. Consequently, the two methods exhibit different sensitivities to population effects and to the line-of-sight extent of the LMC.

At the same time, the LMC hosts large and well characterized populations of RRL \citep{soszynski2019, clementini2023} and T2Cep stars \citep{udalski2018, ripepi2023}, offering a unique opportunity to assess the external validity of population-II calibrations within a common framework. Previous studies have nevertheless reported differences between distance estimates based on various tracers \citep[e.g.][]{bhardwaj2023, sicignano2024}, highlighting the importance of controlling systematic effects at the level of a few hundredths of a magnitude.

In this work, we apply the calibration derived in L26 to a large and homogeneous sample of RRL and T2Cep stars in the LMC from GDR3. This provides a direct external validation of a \gaia-anchored population-II calibration on a key anchor galaxy of the distance network. We derive the distance to the LMC using different classes of pulsating variables within a common framework, assess the consistency between tracers, and compare our results with independent distance determinations. In addition, we exploit the individual stellar distances to investigate the three-dimensional structure of the LMC and to constrain its global disk geometry.

The paper is structured as follows. Section~\ref{sec:dataset} presents the dataset and sample selection. Section~\ref{sec:methodology} describes the methodology, including the distance determination and the geometric modeling. Section~\ref{sec:results} presents the results for the LMC distance and structure. Finally, Section~\ref{sec:conclusions} summarizes our conclusions.

\section{Dataset}\label{sec:dataset}
This analysis is based on astrometric and photometric observations obtained by the European Space Agency’s \gaia\ mission \citep{gaia2016}. We used data from GDR3 \citep{gaiaDR32023}, which provides homogeneous all-sky parallaxes, proper motions, and multi-band photometry, together with a dedicated catalog of variable stars. In particular, GDR3 delivers precise mean magnitudes in the $G$, $BP$, and $RP$ bands, as well as light-curve parameters and variability classifications for large samples of RRL and T2Cep across the LMC. These measurements enabled a consistent application of the absolute Leavitt-law calibration derived in L26 to an external galaxy.

\subsection{RR Lyrae and Type II Cepheid Sample}\label{sec:rrl_t2cep_sample}
We extracted RRL and T2Cep candidates from the GDR3 variability catalogs \citep{gaiaDR32023}. We considered stars classified as RR Lyrae pulsating in the fundamental mode or first overtone, as well as type II Cepheids, using the variability classifications provided in the GDR3 tables. Following L26, we retained only sources processed by the Specific Object Study (SOS) pipeline, which provides homogeneous Fourier light-curve parameters and intensity-averaged photometry optimized for pulsating variables \citep{ripepi2019, clementini2023}. This selection ensured consistency with the calibration relations adopted from L26.

\begin{figure}
    \centering
    \includegraphics[width=\hsize]{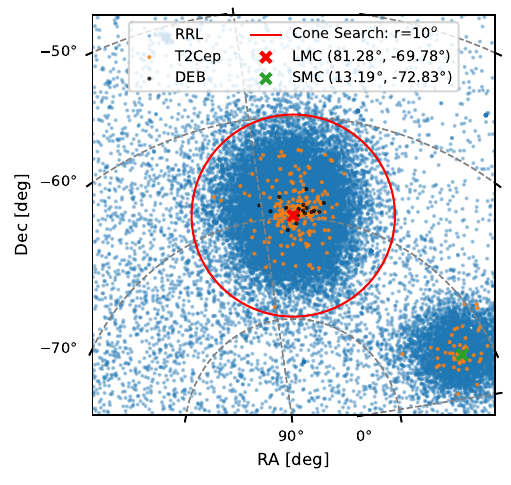}
    \caption{Spatial distribution of \gaia\ DR3 RRL and T2Cep in the vicinity of the LMC, shown in equatorial coordinates. The red circle indicates the $10^\circ$-radius cone selection centered on $(\alpha, \delta) = (81.28^\circ, -69.78^\circ)$. The LMC and SMC centers are marked by red and green crosses, respectively. Detached eclipsing binaries used as independent distance indicators by \citet{pietrzyski2019} are also shown. The selected region encompasses both the main body and the extended halo of the LMC.}
    \label{fig:LMC_cone_search}
\end{figure}

We first defined a broad spatial selection encompassing the LMC. Sources were retained within a circular region of radius $10^\circ$ centered on $(\alpha, \delta) = (81.28^\circ, -69.78^\circ)$, corresponding to the adopted optical center of the LMC \citep{gaiageometry2021, vandermarel2001}. The chosen radius was sufficiently large to include both the main body and the extended halo traced by the old stellar population, while allowing for subsequent kinematic and photometric filtering of residual Milky Way foreground contaminants. Although \gaia\ variability classifications are known to include contamination in crowded regions such as the LMC, particularly at the level of the initial candidate selection, the combination of SOS processing, the quality cuts described below, and the subsequent outlier rejection procedure substantially mitigated these effects. This initial cone selection yielded 34\,899 RRL and 218 T2Cep prior to additional cleaning, and is illustrated in Fig.~\ref{fig:LMC_cone_search}.

The progressive reduction of the sample after each filtering step is summarized in Table~\ref{tab:selection}. Starting from this initial sample, we applied a sequence of kinematic and photometric quality criteria to isolate a clean LMC population. Following the quality-cut strategy adopted by \citet{anderson2024}, we implemented filters designed to suppress Milky Way foreground contamination and remove sources with unreliable \gaia\ photometry or astrometry.

We first selected stars consistent with the systemic motion of the LMC using an elliptical cut in the $(\mu_{\alpha*}, \mu_\delta)$ plane:
\begin{equation}
\left( \frac{\mu_{\alpha*} - 1.83}{1.59} \right)^2 +
\left( \frac{\mu_{\delta} - 0.20}{2.30} \right)^2 < 1,
\end{equation}
where proper motions are expressed in mas\,yr$^{-1}$. The adopted center and dispersions correspond to the bulk motion and intrinsic spread of the LMC population.

To further reject nearby foreground stars, we excluded sources satisfying
\begin{equation}
\frac{\varpi}{\sigma_{\varpi}} > 5
\quad \text{and} \quad
\varpi > 0.1 \ \mathrm{mas},
\end{equation}
where $\varpi$ denotes the parallax corrected following \citet{lindegren2021}. These kinematic cuts efficiently removed stars with significant positive parallaxes inconsistent with LMC distances.

We then imposed photometric quality criteria to ensure reliable mean magnitudes. Following L26, we adopted the same photometric filtering strategy used for the calibration sample in order to maintain full consistency between the calibration relations and the present analysis. Specifically, we required
\begin{align}
\texttt{ipd\_frac\_multi\_peak} &< 10, \\
\texttt{num\_clean\_epochs\_g} &> 10, \\
\texttt{num\_clean\_epochs\_bp} &> 10, \\
\texttt{num\_clean\_epochs\_rp} &> 10,
\end{align}
and removed sources lacking valid intensity-averaged magnitudes in the $G$, $BP$, or $RP$ bands.

To mitigate blending and crowding effects, we adopted the \gaia\ DR3 photometric quality diagnostics \citep[e.g.,][]{riello2021}. We required
\begin{equation}
\beta < 0.99,
\end{equation}
where $\beta$ is the fraction of blended $BP$ and $RP$ transits, and additionally imposed
\begin{equation}
\texttt{ipd\_frac\_odd\_win} < 7,
\qquad
\mathrm{RUWE} < 1.4.
\end{equation}
An additional color-consistency requirement was applied using the \gaia\ $C^\star$ parameter \citep{riello2021},
\begin{equation}
|C^\star| < 3\,\sigma_{C^\star},
\end{equation}
to exclude sources with anomalous colors indicative of problematic flux calibration.

Finally, variability-specific constraints were imposed. For RRL, we required the availability of reliable Fourier parameters, including defined $R_{21}$ values for all stars and defined $\phi_{31}$ values for RRab pulsators. For T2Cep, we excluded RV~Tau variables by removing stars with periods exceeding 20\,days \citep{cruzreyes2025}.

After all selection criteria, the final working sample comprised 17\,160 RR~Lyrae stars and 121 type~II Cepheids distributed across the LMC.
\begin{table}
\caption{Sample reduction after successive filtering steps. Numbers in parentheses indicate the number of stars removed at each step.}
\label{tab:selection}
\centering
\begin{tabular}{lcc}
\hline\hline
Selection step & RR Lyrae & T2Cep \\
\hline
Initial cone selection & 34\,899 & 218 \\
Proper motion selection & 32\,027 (-2\,872) & 215 (-3) \\
Foreground parallax cut & 32\,010 (-17) & 215 (0) \\
Photometric quality cuts & 27\,979 (-4\,031) & 206 (-9) \\
Remove NaN photometry & 27\,673 (-306) & 205 (-1) \\
Blending / RUWE cuts & 24\,783 (-2\,890) & 166 (-39) \\
$C^\star$ color-excess cut & 22\,527 (-2\,256) & 147 (-19) \\
Fourier parameter cuts & 17\,160 (-5\,367)$^{a}$ & -- \\
Remove RV Tau ($P > 20$ d) & -- & 121 (-26) \\
\hline
Final sample & \textbf{17\,160} & \textbf{121} \\
\hline
\end{tabular}
\tablefoot{
\tablefoottext{a}{Includes requirements on the availability of $R_{21}$ for all RRL and $\phi_{31}$ for RRab, as these Fourier parameters are required to estimate photometric metallicities.}
}
\end{table}

\subsection{Wesenheit magnitudes} \label{sec:wesenheit}

We adopted extinction-insensitive Wesenheit magnitudes from the GDR3 intensity-averaged mean magnitudes, using exactly the same formulation and definition as in L26 to ensure full consistency with the absolute calibration.

The \gaia\ Wesenheit magnitude is defined as
\begin{equation}
m_G^{W} = m_G - R_G^{W}\,(m_{G_{\mathrm{BP}}} - m_{G_{\mathrm{RP}}}),
\end{equation}
following the original Wesenheit formalism introduced by \citet{madore1982} and its implementation for \gaia\ pulsating variables \citep[e.g.,][]{ripepi2019}. We adopted a fixed color coefficient of $R_G^{W} = 1.867$ for both RRL and T2Cep, following \citet{cruzreyes2024}.

\subsection{Metallicity}\label{sec:metallicity}
We adopted the same metallicity treatment as in L26. In that work, the metallicity dependence of the Wesenheit Leavitt laws was found to be small, and variants neglecting metallicity effects altogether yielded distance calibrations consistent with the baseline solution within uncertainties. For consistency with the adopted calibration, we nevertheless retained the same formalism here. For RRL, photometric metallicities were derived using the relations of \citet{li2023} based on \gaia\ Fourier light-curve parameters, ensuring full consistency with the calibration sample. The following $P$--$\phi_{31}$--$R_{21}$--[Fe/H] relations were applied for RRab and RRc, respectively:
\begin{align}
    \begin{split}
        [\textrm{Fe/H}]_{RRab} &= -1.888 - 5.772\cdot(P - 0.6) \\
        & \quad + 1.065\cdot(R_{21} - 0.45) + 1.090\cdot(\phi_{31} - 2)
    \end{split} \label{RRab_Li}\\
    \begin{split}
        [\textrm{Fe/H}]_{RRc} &= -1.737-9.968\cdot(P-0.3) \\
        & \quad  -5.041\cdot(R_{21}-0.20).
    \end{split} \label{RRc_Li}
\end{align}

For T2Cep, we adopted a representative metallicity of $\mathrm{[Fe/H]} = -1.5$, corresponding to the mean metallicity of the old stellar population of the LMC.

\section{Methodology}\label{sec:methodology}
\subsection{Adopted Leavitt Laws}\label{sec:PLR}
Individual distances to RRL and T2Cep variables were derived using the absolute PLR calibrated in L26. In the present work, these relations were adopted without modification, and all coefficients were kept fixed to their published calibration values. This was an important aspect of the analysis because the LMC differs substantially from the Galactic GC environment used for the calibration: it exhibits a significant line-of-sight extent and generally lower photometric quality due to crowding and observational conditions. Adopting the GC calibration therefore allowed us to investigate both the global distance and the three-dimensional structure of the LMC using a homogeneous externally calibrated distance scale. The relations are expressed in terms of the GDR3 Wesenheit magnitude defined in Sect.~\ref{sec:wesenheit}. The absolute magnitudes are given by:
\begin{align}
M_{\mathrm{RRab}} &= (-0.300 \pm 0.020) \notag \\
&+ (-2.582 \pm 0.047)\left(\log P + 0.25\right) \notag \\
&+ (-0.028 \pm 0.008)\left([\mathrm{Fe/H}] + 1.5\right) ,
\label{eq:PLR_RRab} \\
M_{\mathrm{RRc}} &= (0.004 \pm 0.021) \notag \\
&+ (-2.497 \pm 0.095)\left(\log P + 0.5\right) \notag \\
&+ (0.013 \pm 0.010)\left([\mathrm{Fe/H}] + 1.5\right) ,\, \mathrm{and}
\label{eq:PLR_RRc} \\
M_{\mathrm{T2Cep}} &= (-3.407 \pm 0.027) \notag \\
&+ (-2.506 \pm 0.039)\left(\log P - 1\right) \notag \\
&+ (0.107 \pm 0.043)\left([\mathrm{Fe/H}] + 1.5\right).
\label{eq:PLR_T2Cep}
\end{align}

The period and metallicity terms are written relative to pivot values in order to minimize correlations between the fitted coefficients in the calibration analysis. The quoted uncertainties correspond to the statistical uncertainties of the calibration coefficients. The uncertainties on the zero point and slope coefficients represent global calibration uncertainties common to the entire sample. They were therefore not propagated on a star-by-star basis, but were treated as correlated systematic contributions to the final LMC distance modulus (see Sect.~\ref{sec:uncertainties}).

\subsection{Distance modulus estimation}\label{sec:distances}
We estimated the distance modulus of the LMC by first computing an individual distance modulus for each variable star from its GDR3 Wesenheit magnitude and the calibrated PLR relations (Sect.~\ref{sec:PLR}). 
For a star $i$, the model can be written as
\begin{equation}
m_i = \mu_{\mathrm{LMC}} + M_i(\log P_i,\mathrm{[Fe/H]}_i) + \varepsilon_i,
\end{equation}
where $m_i$ is the observed magnitude, $\mu_{\mathrm{LMC}}$ is the unknown LMC distance modulus, $M_i$ is the absolute magnitude predicted by the appropriate PLR for the variability class (Eqs.~\ref{eq:PLR_RRab}--\ref{eq:PLR_T2Cep}), and $\varepsilon_i$ represents random scatter.

We defined the individual distance modulus estimate as
\begin{equation} \label{eq:mu_i}
\mu_i \equiv m_i - M_i(\log P_i,\mathrm{[Fe/H]}_i).
\end{equation}
Under the assumption of Gaussian errors, each $\mu_i$ is modeled as
\begin{equation}
\mu_i = \mu_{\mathrm{LMC}} + \epsilon_i,
\end{equation}
with variance
\begin{equation}
\sigma_i^2 \equiv \sigma_{\mu_i}^2
= \sigma_{m_i}^2 + \sigma_{\mathrm{extra}}^2 ,
\end{equation}
where $\sigma_{m_i}$ is the photometric uncertainty on the Wesenheit magnitude and $\sigma_{\mathrm{extra}} = 0.22$\,mag represents the observed dispersion around the adopted PLR. In Galactic GCs, where line-of-sight depth effects are negligible, the intrinsic scatter of the calibrated relations is significantly smaller than in the LMC. The larger dispersion observed in the LMC likely reflects additional contributions from its geometric depth, residual metallicity effects, evolutionary differences, and remaining photometric uncertainties. We therefore adopted an additional scatter term matched to the observed RRab dispersion in the LMC (see Sect.~\ref{sec:baseline_rrab_distance}) in order to obtain realistic weights for individual distance estimates. The likelihood for $\mu_{\mathrm{LMC}}$ is then
\begin{equation}
\mathcal{L}(\mu_{\mathrm{LMC}}) = \prod_i 
\frac{1}{\sqrt{2\pi\sigma_i^2}}
\exp\!\left[-\frac{(\mu_i-\mu_{\mathrm{LMC}})^2}{2\sigma_i^2}\right],
\end{equation}
and the corresponding log-likelihood is
\begin{equation}
\ln \mathcal{L} = -\frac{1}{2}\sum_i\left[
\frac{(\mu_i-\mu_{\mathrm{LMC}})^2}{\sigma_i^2}
+ \ln(2\pi\sigma_i^2)
\right].
\end{equation}
Maximizing this likelihood yields the standard inverse-variance weighted mean:
\begin{equation} \label{eq:mu_LMC}
\hat{\mu}_{\mathrm{LMC}} = \frac{\sum_i w_i \mu_i}{\sum_i w_i},
\qquad
w_i = \frac{1}{\sigma_i^2},
\end{equation}
with statistical uncertainty
\begin{equation}
\sigma_{\hat{\mu},\mathrm{stat}} = \left(\sum_i w_i\right)^{-1/2}.
\end{equation}

Uncertainties on the calibration coefficients of the adopted PLR represent global correlated calibration uncertainties and were therefore not included in the star-by-star variances $\sigma_i^2$. They were added as systematic contributions to the final uncertainty budget on $\hat{\mu}_{\mathrm{LMC}}$ (Sect.~\ref{sec:uncertainties}).

\subsection{Uncertainty budget}\label{sec:uncertainties}
In addition to the statistical component discussed above, the inferred LMC distance modulus depends on the calibration parameters of the adopted Leavitt laws. 
For RRab variables the relation can be written as
\begin{equation}
    M = \alpha_0 + \beta_0\,(\log P - \log P_0) +
        \alpha_m\,([\mathrm{Fe/H}] - [\mathrm{Fe/H}]_0),
\end{equation}
with analogous expressions for RRc and T2Cep variables, expressed relative to the RRab zero point through offsets $\Delta\alpha$. 
Uncertainties in these calibration parameters therefore propagate into the final estimate of the LMC distance modulus.

To account for this effect, we propagated the calibration uncertainties using the full covariance matrix of the calibration parameters. 
Denoting by $\boldsymbol{\theta}$ the vector of calibration parameters and by $\mathbf{C}_\theta$ their covariance matrix, the variance on the LMC distance modulus induced by calibration uncertainties is obtained through linear error propagation,
\begin{equation}
    \sigma_{\mathrm{syst}}^2 =
    \mathbf{g}^{\mathsf T}\,\mathbf{C}_\theta\,\mathbf{g},
\end{equation}
where
\begin{equation}
    \mathbf{g} =
    \frac{\partial \mu_{\mathrm{LMC}}}{\partial \boldsymbol{\theta}}
\end{equation}
is the gradient of the final distance modulus with respect to the calibration parameters (see Eqs.~\ref{eq:mu_i} and \ref{eq:mu_LMC}). 
Since the LMC distance modulus is computed as the weighted mean of the individual stellar moduli, the components of $\mathbf{g}$ correspond to weighted sums of the derivatives of $\mu_i$ with respect to each calibration parameter.

This formulation naturally accounts for the correlated uncertainties between the calibration parameters and treats the calibration uncertainty as a global systematic affecting all stars simultaneously, rather than as an independent contribution for each object. The total uncertainty on the LMC distance modulus is therefore expressed as the quadratic sum of the statistical and calibration contributions,
\begin{equation}
    \sigma_{\mathrm{tot}}^2 =
    \sigma_{\mathrm{stat}}^2 + \sigma_{\mathrm{syst}}^2 .
\end{equation}

\subsection{Outlier rejection}\label{sec:outliers}
Outliers were identified following the same iterative procedure adopted in L26. Chauvenet’s criterion was applied separately to each variability class (RRab, RRc, and T2Cep) using the residuals of the individual distance moduli relative to the current estimate of the mean LMC distance modulus.

The algorithm proceeded iteratively. At each step, the star with the largest absolute normalized residual was identified. If it did not satisfy Chauvenet’s criterion, it was removed from the sample, and the weighted mean distance modulus was recomputed. The procedure was repeated until all remaining stars satisfied the criterion.

\subsection{Three-dimensional coordinate system}\label{sec:coordinates}
To investigate the three-dimensional structure of the LMC, the sky coordinates and distances of the variable stars were converted into a Cartesian coordinate system centered on the LMC. We adopted the formalism introduced by \citet{vandermarel&cioni2001}, which is commonly used in studies of the LMC geometry and has been applied in several previous works \citep[e.g.,][]{weinberg2001, cusano2021, deb&singh2018}. The projected Cartesian coordinates are given by
\begin{align}
x &= - D \cos\delta \sin(\alpha - \alpha_0), \notag \\
y &= D \left[\sin\delta \cos\delta_0 - \cos\delta \sin\delta_0 \cos(\alpha-\alpha_0)\right], \notag \\
z &= D_0 - D \left[\sin\delta \sin\delta_0 + \cos\delta \cos\delta_0 \cos(\alpha-\alpha_0)\right],
\end{align}
where $(\alpha,\delta)$ denote the equatorial coordinates (right ascension and declination) of a star, $(\alpha_0,\delta_0)$ the adopted center of the LMC, $D$ the distance to the star, and $D_0$ the distance to the LMC center.

The distances $D$ used in the coordinate transformation were derived from the individual distance moduli obtained in Sect.~\ref{sec:distances}. For each star, the distance is computed as
\begin{equation}
D = 10^{(\mu_i + 5)/5}\ \mathrm{pc},
\end{equation}
where $\mu_i$ is the individual distance modulus inferred from the adopted Leavitt laws (Eq.~\ref{eq:mu_i}). These distances provide the line-of-sight coordinate necessary to reconstruct the three-dimensional spatial distribution of the sample.

In this coordinate system, the origin corresponds to the adopted center of the RRab distribution. The $x$-axis points toward decreasing right ascension, approximately west on the sky, the $y$-axis toward increasing declination, north, and the $z$-axis is directed along the line of sight toward the observer. Rather than adopting a fixed literature center, we defined $(\alpha_0,\delta_0,D_0)$ using the barycenter of the RRab sample itself, ensuring consistency between the coordinate system and the stellar population used to determine the LMC distance and geometry. Alternative center definitions, such as the optical center from \citet{vandermarel2001}, were also considered and were found to have a negligible impact on the inferred geometry, with variations well below the formal uncertainties of the fit. The fitted LMC geometry was subsequently used to derive differential geometric distance corrections between different reference positions within the galaxy.

\subsection{Disk geometry and plane fitting}\label{sec:plane_fit}
To characterize the large-scale geometry of the LMC, we modeled the spatial distribution of the stars using a planar approximation. On kiloparsec scales, the stellar disk of the LMC can be described, to first order, as a tilted plane, allowing its global orientation to be quantified in terms of an inclination and a position angle of the line of nodes. Following previous studies of the LMC structure \citep[e.g.,][]{vandermarel&cioni2001, weinberg2001, deb&singh2018, cusano2021}, we therefore adopted a planar model to describe the three-dimensional distribution of the variable stars.

In the Cartesian coordinate system defined in Sect.~\ref{sec:coordinates}, the disk is modeled by a plane of the form
\begin{equation}
z = ax + by + c ,
\end{equation}
where $a$, $b$, and $c$ are the parameters of the plane model. The plane parameters were determined through a weighted least-squares fit, using the individual statistical distance uncertainties as weights. The global calibration systematic uncertainty was not included in the weighting, since it affects all stars coherently and therefore does not contribute to the relative geometry of the LMC. To mitigate the impact of outliers, the fit was iteratively refined using sigma-clipping on the residuals, and the final solution was obtained after convergence. Uncertainties on the fitted parameters were estimated via bootstrap resampling of the dataset.

The inclination of the disk relative to the plane of the sky is derived from the normal vector to the plane and is given by
\begin{equation}
i = \arctan\left(\sqrt{a^2 + b^2}\right),
\end{equation}
where $i = 0^\circ$ corresponds to a face-on configuration.

The position angle of the line of nodes, defined as the intersection between the disk plane and the plane of the sky, is obtained from
\begin{equation}
\Theta = \mathrm{atan2}(-b, a) ,
\end{equation}
where $\Theta$ is measured counterclockwise from North toward East and is defined modulo $180^\circ$. This convention explicitly accounts for the adopted coordinate system in which $x$ increases toward the west.

The fitted plane therefore provides a direct estimate of the global orientation of the LMC disk, allowing both its inclination and the position angle of the line of nodes to be derived from the three-dimensional distribution of the RRab stars. In addition, the fitted geometry provides a simple framework to compute geometric distance corrections between different adopted center definitions or positions within the LMC.

An alternative approach consists in determining the principal axes of the three-dimensional stellar distribution via the inertia tensor, effectively modeling the system as a triaxial ellipsoid \citep[e.g.,][]{deb&singh2018, cusano2021}. However, this method is primarily sensitive to the overall shape of the distribution and can be significantly affected by the finite vertical thickness of the tracer population as well as by distance uncertainties. For the present dataset, such an approach did not yield a stable estimate of the disk inclination. Since our goal was to recover the orientation of the mean LMC disk, we therefore adopted a planar model, which more directly traces the large-scale disk geometry.

\section{Results}\label{sec:results}
We present the results obtained from applying the calibrated PLRs to the LMC variable-star sample described in Sect.~\ref{sec:dataset}. The large number of stars enabled both a precise determination of the weighted mean LMC distance modulus and an investigation of the three-dimensional structure of the system. We first derived the weighted mean LMC distances obtained from the different classes of variable stars and assessed the robustness of these measurements (Sect.~\ref{sec:lmc_distance}). We then used the individual stellar distances to study the spatial distribution and geometry of the LMC (Sect.~\ref{sec:lmc_geometry}). The resulting geometric model furthermore allowed geometric distance corrections to be computed between different adopted center definitions within the galaxy.

\subsection{LMC distance determination}\label{sec:lmc_distance}
The distance to the LMC was derived from the individual stellar distance moduli obtained using the calibrated PL relations described in Sect.~\ref{sec:methodology}. Our sample contains several classes of pulsating variables, including RRab, RRc, and T2Cep, which can be used as complementary distance tracers within the same calibration framework. In this work, we adopted the RRab variables as the baseline tracer for the distance determination for several reasons.

First, the calibration of the PL relations adopted in this work is anchored on the RRab population. In the joint calibration presented in L26, the RRab PLR defines the absolute zero point, while the relations for RRc and T2Cep are expressed relative to this reference through additive offsets. The distance determination based on RRab variables therefore corresponds to the most direct application of the calibration, whereas the use of other tracers relies on additional relative parameters and uncertainties that may introduce further systematic uncertainties.

Second, RRab constitute the dominant pulsating population in the LMC sample. The present dataset contains $N_{\mathrm{RRab}} = 12532$, compared to $N_{\mathrm{RRc}} = 4628$ and $N_{\mathrm{T2Cep}} = 121$. The large size of the RRab sample therefore provides the most robust statistical estimate of the mean distance modulus. In terms of photometry, RRab and RRc occupy a similar magnitude range in the \gaia\ bands used to construct the Wesenheit magnitudes, with median magnitudes of $(G, BP, RP) = (19.24, 19.42, 18.82)$\,mag for RRab and $(19.25, 19.41, 18.92)$\,mag for RRc. RRab are nevertheless slightly brighter on average in the $RP$ band and remain the best-constrained population overall due to their significantly larger sample size and their role as the reference calibration population in L26.

Third, RRab provide the most robust tracer from the standpoint of classification and population transfer. Their asymmetric light curves make them easier to identify than the more sinusoidal RRc variables, reducing the risk of contamination from eclipsing binaries and other contaminants. In addition, the calibration sample used to derive the PLRs in L26 consists of GCs, which host predominantly old and chemically relatively homogeneous stellar populations. By contrast, the LMC contains a more complex stellar environment with broader age and chemical distributions. For example, \citet{sarbadhicary2021} showed that the LMC RRL population is not exclusively ancient but likely includes a contribution from intermediate-age progenitors. Population-dependent effects may therefore affect the transferability of the relative calibration between pulsation modes derived from the GC sample. Similar environmental differences have also been reported for T2Cep populations, whose period distributions differ between GCs and the more heterogeneous environment of the LMC \citep{cruzreyes2025}.

We found evidence for such an effect in the relative zero points of the RRab and RRc relations. In L26, the relative calibration between RRab and RRc was determined self-consistently from the GC sample. In the LMC, however, applying the same calibration led to a systematic offset of $\sim0.05$--$0.07$\,mag between the RRab and RRc intercepts. This discrepancy is more naturally interpreted as a population-dependent difference between the LMC and GC samples than as a failure of the absolute RRab calibration. We therefore adopted RRab as the baseline tracer, while using RRc and T2Cep primarily as consistency checks.

Finally, the large number of RRab already allowed the uncertainty on the mean distance modulus to reach the floor set by the intrinsic scatter of the PL relations. Using RRab alone, we obtained a statistical uncertainty of 0.0024\,mag and a systematic calibration uncertainty of 0.0202\,mag, while combining all tracers (RRab+RRc+T2Cep) yielded nearly identical values of 0.0020\,mag and 0.0201\,mag, respectively. The negligible gain in precision demonstrates that including additional tracers does not significantly improve the overall uncertainty of the distance measurement.

\subsubsection{Baseline RRab distance} \label{sec:baseline_rrab_distance}
\begin{figure*}
    \centering
    \includegraphics[width=\hsize]{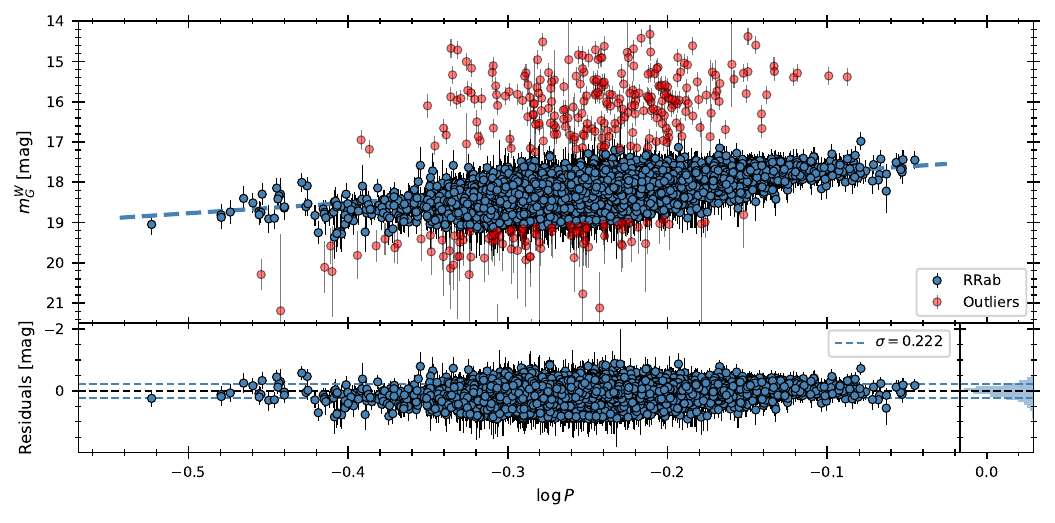}
    \caption{PLR of RRab stars in the LMC. Blue points show the retained sample, red points indicate rejected outliers, and the dashed line corresponds to the calibrated RRab PLR shifted to the RRab distance modulus, $\mu_{\mathrm{LMC}} = 18.423 \pm 0.002\,(\mathrm{stat}) \pm 0.020\,(\mathrm{syst})$\,mag. The lower panel shows the residuals, with dashed lines indicating the $1\sigma$ dispersion.}
    \label{fig:RRab_LL}
\end{figure*}

Using the RRab sample, we derived a distance modulus of $\mu_{\mathrm{LMC}} = 18.423 \pm 0.002\,(\mathrm{stat}) \pm 0.020\,(\mathrm{syst})$\,mag. This value was obtained from the inverse-variance weighted mean of the individual stellar distance moduli, following the procedure described in Sect.~\ref{sec:distances}. An iterative outlier rejection based on Chauvenet’s criterion was applied, resulting in a final sample of 12\,193 RRab stars, with 339 objects rejected as outliers. The statistical uncertainty ($\sigma_{\mathrm{stat}} = 0.0024$\,mag) reflects the uncertainty on the mean distance modulus, while the systematic term ($\sigma_{\mathrm{syst}} = 0.0202$\,mag) accounts for the propagation of the PLR calibration uncertainties. Figure~\ref{fig:RRab_LL} illustrates the final fit and the distribution of retained and rejected sources.

The distance modulus derived from RRab stars is consistent with recent high-precision determinations of the LMC distance. In particular, it can be compared to the DEB measurement of $\mu_{\mathrm{LMC}} = 18.477 \pm 0.026$\,mag reported by \citet{pietrzyski2019}. However, the two measurements do not refer exactly to the same location within the LMC. Our RRab distance corresponds to the adopted RRab barycenter, whereas the DEB distance is tied to a different geometric center of the galaxy. Using the fitted disk geometry derived in Sect.~\ref{sec:lmc_geometry}, we computed the corresponding geometric correction between the two center definitions and found it to be negligible compared to the current uncertainties (see Sect.~\ref{sec:distance_corrections}). The two values differ by $\Delta \mu = 0.054$\,mag, corresponding to a $\sim1.7\sigma$ offset when accounting for the combined uncertainties. While our RRab-based estimate is slightly smaller than the DEB value, the two measurements remain statistically compatible.

A key consistency check is provided by the recent work of \citet{cruzreyes2026}, who derived the LMC distance using the TRGB calibrated from the same set of GCs anchored to GDR3 parallaxes and based on the same underlying distance scale from L26. Their result, $\mu_{\mathrm{LMC}} = 18.447^{+0.036}_{-0.042}$\,mag, differs from our RRab-based estimate by only $\sim0.024$\,mag, well within the combined uncertainties. This agreement demonstrates that independent Population-II distance indicators, based on fundamentally different observables, yield consistent distances when tied to a common geometric calibration framework.

\subsubsection{Impact of additional tracers}
We quantified the effect of using different classes of variable stars on the inferred LMC distance modulus. Distances were derived separately from RRc and T2Cep, as well as from the combined sample, in order to characterize systematic differences between tracers. The results are summarized in Table~\ref{tab:lmc_variants}, together with the sky-position barycenters of the corresponding samples.

Using RRc alone, we obtained a distance modulus of $\mu_{\mathrm{LMC}} = 18.368 \pm 0.004\,(\mathrm{stat}) \pm 0.020\,(\mathrm{syst})$\,mag, which is lower than the RRab-based value by $\sim0.055$\,mag. The distance derived from T2Cep is $\mu_{\mathrm{LMC}} = 18.405 \pm 0.022\,(\mathrm{stat}) \pm 0.025\,(\mathrm{syst})$\,mag, although the small sample size and larger calibration uncertainties limit its constraining power. Combining all tracers yields $\mu_{\mathrm{LMC}} = 18.407 \pm 0.002\,(\mathrm{stat}) \pm 0.020\,(\mathrm{syst})$\,mag, which is slightly lower than the RRab-only result but remains consistent within uncertainties. The combined fit is shown in Fig.~\ref{fig:all_LL} in Appendix~\ref{app:additional_figs}.

The different tracer populations do not sample exactly the same regions of the LMC and therefore exhibit slightly different barycenters on the sky (Table~\ref{tab:lmc_variants}). Given the finite geometric extent and inclination of the LMC, part of the observed offsets between tracers could therefore arise from geometric effects. However, using the disk model derived in Sect.~\ref{sec:lmc_geometry}, we found that the corresponding geometric corrections between the different barycenters are negligible compared to the observed $\sim0.05$\,mag RRab--RRc offset. The comparison between tracers therefore reveals a genuine systematic difference between the RRab and RRc populations, consistent with the behavior discussed in Sect.~\ref{sec:lmc_distance}. The inclusion of additional tracers thus introduces small systematic shifts in the inferred distance modulus without providing a significant gain in precision. This behavior likely reflects differences in stellar populations and calibration transferability rather than statistical limitations. These results further motivate the use of RRab as the baseline tracer for the LMC distance determination.

\begin{table}
\caption{Distance moduli derived from different tracers together with the corresponding sky-position barycenters of the samples.}
\label{tab:lmc_variants}
\centering
\begin{tabular}{lcccc}
\hline\hline
Candle & $\mu_{\mathrm{LMC}}$ & $\sigma_{\mathrm{syst}}$ & $\sigma_{\mathrm{stat}}$ & Barycenter $(\alpha,\delta)$ \\
\hline
RRab   & 18.423 & 0.0202 & 0.0024 & $(81.01^\circ,\,-69.42^\circ)$ \\
RRc    & 18.368 & 0.0204 & 0.0037 & $(81.23^\circ,\,-69.30^\circ)$ \\
T2Cep  & 18.405 & 0.0248 & 0.0219 & $(80.85^\circ,\,-69.18^\circ)$ \\
All    & 18.407 & 0.0201 & 0.0020 & $(81.07^\circ,\,-69.39^\circ)$ \\
\hline
\end{tabular}
\end{table}

\subsection{Three-dimensional structure of the LMC}\label{sec:lmc_geometry}
Using the individual distances (Eq.~\ref{eq:mu_i}) derived from the RRab sample, we investigated the three-dimensional structure of the LMC. The spatial distribution of the stars was analyzed in a Cartesian reference frame centered on the LMC, following the formalism described in Sect.~\ref{sec:coordinates}. 

\subsubsection{Spatial distribution of RRab stars}
\begin{figure}
    \centering
    \includegraphics[width=\hsize]{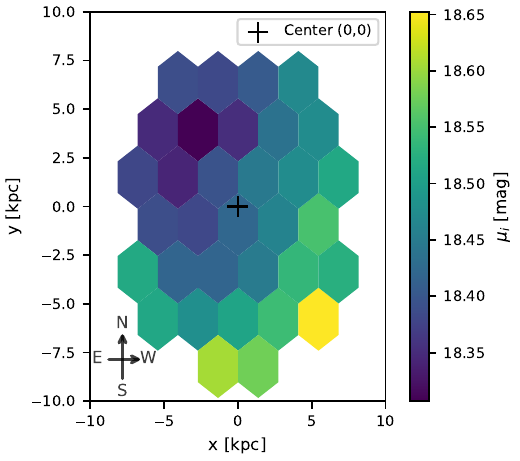}
    \caption{Spatial distribution of RRab stars in the LMC in projected Cartesian coordinates. Each hexagonal bin shows the median distance modulus $\mu_i$ of stars within the bin, with a minimum of 25 stars per cell. A clear large-scale gradient is visible across the LMC. The red cross marks the adopted center of the LMC.}
    \label{fig:Hexbins}
\end{figure}

The spatial distribution of RRab in the LMC is shown in Fig.~\ref{fig:Hexbins}. The map displays the median distance modulus in bins of projected coordinates $(x,y)$, highlighting large-scale variations across the galaxy. A clear gradient is visible, with the north-eastern side of the LMC systematically closer to the observer than the opposite side.

The amplitude of the distance gradient across the field reaches $\sim0.37$\,mag, whereas the residual bin-to-bin scatter around a smooth planar trend is only $\sim0.04$\,mag, indicating a coherent geometric effect rather than random fluctuations. The orientation of this gradient provides a first estimate of the direction of the line of nodes, which is quantified in Sect.~\ref{sec:plane_fit}. A qualitatively similar large-scale gradient is also observed for the RRc population, whereas the number of available T2Cep is insufficient to construct an equivalent spatial map with meaningful resolution.

To further illustrate this effect, Fig.~\ref{fig:Slices} shows the projected spatial distribution of RRab in successive slices of line-of-sight coordinate $z$, each containing an equal number of stars. The barycenter of each slice is indicated by a red cross. A systematic shift of the barycenters is observed from distant (negative $z$) to nearby (positive $z$) slices, confirming that the north-eastern regions of the LMC are closer to the observer. This behaviour reflects the large-scale inclined geometry of the LMC across the surveyed sky area, with different projected regions dominating at different line-of-sight distances.

Our results are consistent with previous studies of the LMC structure. A large-scale distance gradient across the galaxy, interpreted as the signature of an inclined disk, has long been established \citep{vandermarel&cioni2001}. More recent analyses using different stellar tracers have reported a similar global geometry. In particular, studies based on red clump stars and Classical Cepheids have revealed coherent variations in line-of-sight distance across the LMC, with the north-eastern regions lying closer to the observer than the south-western ones \citep{saroon2022,ripepi2022}. Analyses based on RRL also support an inclined and spatially extended structure for the old stellar population \citep{deb&singh2018,cusano2021}. Our RRab distance map is therefore in good agreement with the overall picture of an inclined LMC disk in which the north-eastern side is closer to the observer.

\subsubsection{Planar fit and disk geometry}\label{sec:plane_fit}
\begin{figure*}
    \centering
    \includegraphics[width=\hsize]{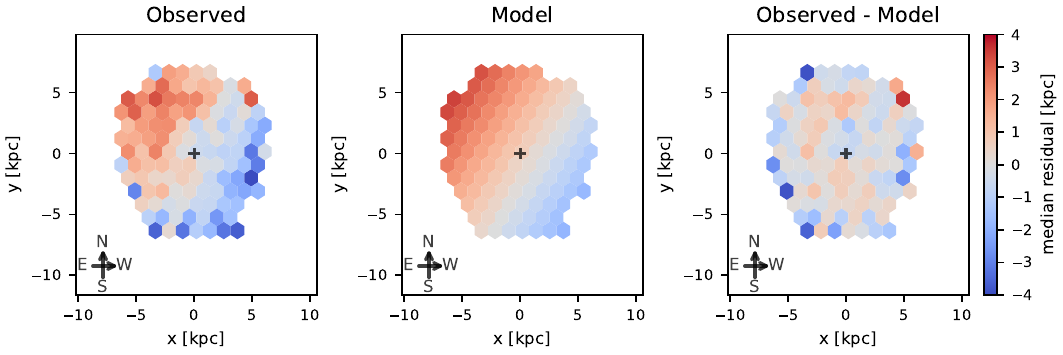}
    \caption{Three-dimensional structure of the LMC traced by RRab stars. \textit{Left:} Median line-of-sight distance $z$ in bins of projected coordinates $(x,y)$. \textit{Middle:} Best-fit planar model. \textit{Right:} Residuals (observed $-$ model). All panels share the same color scale. The fitted plane reproduces the large-scale distance gradient across the galaxy, with the north-eastern regions of the LMC lying closer to the observer. The residual map shows no significant coherent large-scale structure, indicating that a planar model provides an adequate first-order description of the RRab geometry.}
    \label{fig:PlaneFitComparison}
\end{figure*}

The three-dimensional distribution of RRab stars was modeled with a plane of the form $z = ax + by + c$, fitted using a robust, sigma-clipped weighted least-squares approach. From the fitted plane coefficients, we derived an inclination of $i = 21.3 \pm 0.7^\circ$ and a position angle of the line of nodes $\Theta = 145.2 \pm 2.2^\circ$, measured North through East. The fitted orientation indicates that the eastern and northern regions are closer to the observer, implying that the north-eastern side of the LMC is tilted toward us, in agreement with the qualitative behaviour described above. Repeating the analysis using the optical center from \citet{vandermarel2001} yielded nearly identical results, with changes below $0.1^\circ$ in inclination and below $0.6^\circ$ in the line-of-nodes position angle, well below the formal uncertainties of the fit. The fitted plane was therefore used only to characterize the large-scale orientation of the RRab component and to compute differential geometric corrections between reference positions.

The observed variation of $z$ across the projected $(x,y)$ plane reveals a clear large-scale gradient, which is well reproduced by the fitted model (Fig.~\ref{fig:PlaneFitComparison}). The residual map shows no significant large-scale structure, indicating that a planar model provides an adequate first-order description of the RRab geometry. Small residual fluctuations remain present in the outer regions, but their amplitude is limited and does not affect the inferred global orientation.

The robustness of the derived parameters was verified through several tests. Ordinary least-squares and weighted least-squares fits without clipping yielded consistent results ($(i, \Theta) = (22.8^\circ, 143.1^\circ)$ and $(i, \Theta) = (21.1^\circ, 144.6^\circ)$, respectively), demonstrating that the inferred geometry is not driven by the adopted weighting scheme. The use of sigma-clipping ensured that outliers did not bias the solution, while bootstrap resampling provided uncertainty estimates.

The derived inclination and line-of-nodes position angle are consistent with previous measurements based on different stellar tracers. Studies using Classical Cepheids and red clump stars typically find $i \sim 23^\circ$--$26^\circ$ and $\Theta \sim 145^\circ$--$160^\circ$ \citep[e.g.,][]{inno2016,ripepi2022,saroon2022}. Our results fall within these ranges, supporting a coherent picture of the LMC as an inclined disk, while minor differences may reflect the older and more spatially extended nature of the RRab population. Although RRL trace an older and dynamically hotter population, their spatial distribution still reproduces the large-scale disk geometry to first order. The present analysis therefore provides an independent RRab-based characterization of the LMC orientation that is fully consistent with the adopted RRab distance scale and can be used to compute differential geometric corrections between different reference positions within the galaxy.

Studies based on RR Lyrae stars using ellipsoidal models report somewhat different orientation parameters \citep{deb&singh2018, cusano2021}. These differences likely arise from the distinct geometric assumptions, as such analyses attempt to model the full three-dimensional structure of the old stellar population rather than the large-scale disk orientation traced by the present approach. We explored more complex ellipsoidal descriptions of the RRab distribution, but found the resulting solutions to be unstable and strongly dependent on the adopted parametrization and fitting procedure. Given the substantial line-of-sight scatter of the individual distance estimates and the absence of significant large-scale residual structure after subtraction of the planar model (Fig.~\ref{fig:PlaneFitComparison}), we therefore adopted a planar description as a robust first-order characterization of the global LMC geometry.

\subsubsection{Geometric distance corrections} \label{sec:distance_corrections}
The fitted disk geometry was used to estimate differential geometric corrections between different reference positions within the LMC. Using the planar model derived above, we evaluated the expected distance modulus offsets between the weighted RRab barycenter adopted in this work and several commonly used reference positions, including the optical center from \citet{vandermarel2001}, the DEB reference position from \citet{pietrzyski2019}, and the barycenters of the different tracer populations.

In all cases, the inferred geometric corrections remained very small, with amplitudes below $\sim0.003$\,mag. In particular, the correction between the RRab barycenter and the DEB reference position was only $+0.0025$\,mag, while the correction to the optical center of \citet{vandermarel2001} was $-0.0011$\,mag. Similarly small corrections were obtained for the RRc, T2Cep, and combined-sample barycenters. These values are negligible compared to both the statistical and systematic uncertainties of the present analysis. We therefore conclude that differences in adopted center definitions do not significantly affect the comparison between the various LMC distance estimates discussed in this work.

\section{Summary and conclusions} \label{sec:conclusions}
We determined the distance and three-dimensional structure of the LMC using a homogeneous sample of RRL and T2Cep variables from GDR3, together with the \gaia-based population-II Leavitt-law calibration from L26. The initial selection covered a broad $10^\circ$-radius region around the LMC, providing wide spatial leverage across both the main body and the extended old stellar population. After applying kinematic, photometric, and variability-based quality cuts, we used RRab stars as the baseline tracer because they define the absolute zero point of the calibration, dominate the final sample, and provide the most robust classification among the available pulsating-star populations.

Using the RRab sample, we obtained a weighted mean LMC distance modulus of
$\mu_{\mathrm{LMC}} = 18.423 \pm 0.002\,(\mathrm{stat}) \pm 0.020\,(\mathrm{syst})~\mathrm{mag}$. The uncertainty is dominated by the systematic contribution from the calibration of the adopted Leavitt law, while the statistical uncertainty on the weighted mean is negligible owing to the large number of RRab stars. The RRab-based distance is slightly smaller than the DEB benchmark of \citet{pietrzyski2019}, differing by $\Delta\mu = 0.054$~mag, corresponding to a $\sim1.7\sigma$ offset when the combined uncertainties are considered. The comparison is not affected by differences in adopted reference positions within the LMC, since the geometric correction between the RRab barycenter and the DEB reference position is only $+0.0025$~mag. Our result is also consistent within the uncertainties with the TRGB distance derived by \citet{cruzreyes2026}, which is based on the same underlying \gaia-anchored GC distance scale as L26 but relies on a fundamentally different observable.

Distances inferred from additional tracers provided useful consistency checks, but did not improve the final precision. RRc stars yielded a distance modulus lower than the RRab value by $\sim0.055$~mag, while T2Cep stars gave a value closer to the RRab result but with substantially larger statistical uncertainty due to the small sample size. The different tracers have slightly different sky-position barycenters; however, the corresponding geometric corrections are below $\sim0.003$~mag and are therefore negligible compared to the observed RRab--RRc offset. This indicates that the offset is not caused by spatial sampling of the inclined LMC disk, but more likely reflects population-dependent effects or limitations in the transferability of the relative calibration between GCs and the more complex LMC environment. This result motivates the use of RRab stars as the baseline distance tracer, with RRc and T2Cep stars treated as complementary diagnostics.

The individual RRab distances also allowed us to reconstruct the large-scale three-dimensional structure of the LMC. The distance map revealed a clear gradient across the galaxy, with the north-eastern regions lying closer to the observer. A planar model provided an adequate first-order description of this gradient, yielding an inclination of $i = 21.3 \pm 0.7^\circ$ and a line-of-nodes position angle of $\Theta = 145.2 \pm 2.2^\circ$, measured North through East. These values agree with previous determinations based on classical Cepheids, red clump stars, and RRL populations, confirming that the old RRab component traces the global orientation of the LMC disk to first order. The fitted geometry further provides a self-consistent framework for computing differential distance corrections between reference positions within the LMC.

Overall, this work demonstrates that \gaia-calibrated population-II pulsators provide a precise and internally consistent route to measure distances in the Local Group and to study the internal structure of nearby galaxies. The agreement between RRab and TRGB distances tied to the same GC-based geometric scale strengthens the role of population-II indicators as a complementary component of the distance network. At the same time, the residual offset between RRab and RRc in the LMC highlights the importance of testing the transferability of relative calibrations across different stellar environments. Future \gaia\ data releases, improved light-curve classifications, and more homogeneous metallicity information should further reduce the dominant systematic uncertainties and improve the accuracy of population-II distance measurements.

\section*{Data availability}
This work is based on publicly available data from \gaia\ Data Release 3, accessible through the \gaia\ Archive. 

\begin{acknowledgements}
This research has received support from the European Research Council (ERC) under the European Union's Horizon 2020 research and innovation programme (Grant Agreement No. 947660). RIA was funded by a Swiss National Science Foundation Eccellenza Professorial Fellowship (award PCEFP2\_194638). Funded by zukunft.niedersachsen, the joint science funding program of the Lower Saxony Ministry of Science and Culture and the Volkswagen Foundation.

This work has made use of data from the European Space Agency (ESA) mission {\it Gaia} (\url{https://www.cosmos.esa.int/gaia}), processed by the {\it Gaia} Data Processing and Analysis Consortium (DPAC, \url{https://www.cosmos.esa.int/web/gaia/dpac/consortium}). Funding for the DPAC has been provided by national institutions, in particular the institutions participating in the {\it Gaia} Multilateral Agreement. 

This research has made use of NASA's Astrophysics Data System; the SIMBAD database and the VizieR catalog access tool\footnote{\url{http://cdsweb.u-strasbg.fr/}} provided by CDS, Strasbourg.

This work made use of the following open-source software: 
\texttt{Python}\footnote{\url{https://www.python.org}}, 
\texttt{NumPy} \citep{harris2020array}, 
\texttt{Pandas} \citep{reback2020pandas}, 
\texttt{Matplotlib} \citep{hunter2007matplotlib}, 
\texttt{SciPy} \citep{virtanen2020scipy}, 
\texttt{Astropy} \citep{astropy:2013, astropy:2018, astropy:2022}, 
and \texttt{Astroquery} \citep{ginsburg2019}. 
We are grateful to the developers and maintainers of these packages for making their work available to the community.
\end{acknowledgements}

\bibliographystyle{aa}
\bibliography{refs.bib}

\begin{appendix}
\section{Additional figures} \label{app:additional_figs}
This appendix presents supplementary figures supporting the distance and structural analyses discussed in Sect.~\ref{sec:results}.
\begin{figure*}
    \centering
    \includegraphics[width=\hsize]{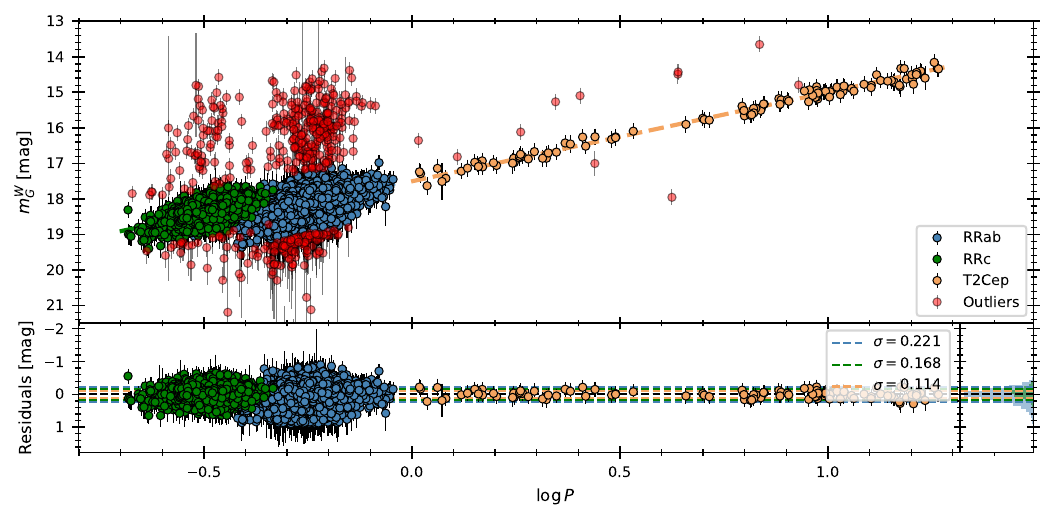}
    \caption{Joint PLR fit to the LMC sample including RRab, RRc, and T2Cep stars. Blue, green, and yellow points show RRab, RRc, and T2Cep stars, respectively, while red points indicate rejected outliers. The dashed lines correspond to the calibrated PLRs shifted to the combined-sample distance modulus, $\mu_{\mathrm{LMC}} = 18.407 \pm 0.002\,(\mathrm{stat}) \pm 0.020\,(\mathrm{syst})$\,mag. The lower panel shows the residuals, with dashed lines indicating the $1\sigma$ dispersion for each tracer.}
    \label{fig:all_LL}
\end{figure*}

\begin{figure*}
    \centering
    \includegraphics[width=\hsize]{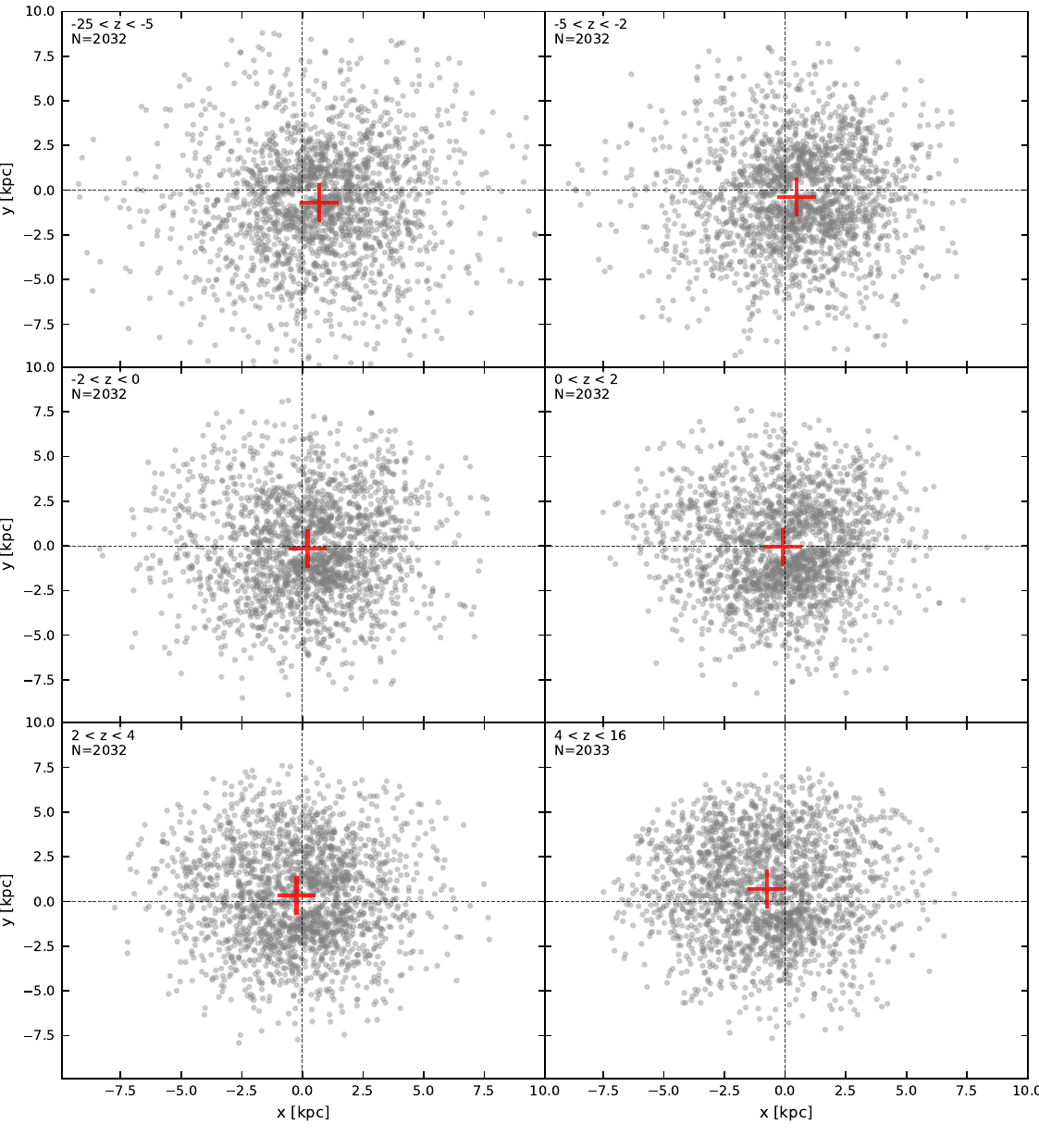}
    \caption{Projected spatial distribution of RRab in successive slices of line-of-sight coordinate $z$, each containing an equal number of stars. Gray points show individual stars, while red crosses indicate the barycenter of each slice. Positive $z$ values correspond to stars closer to the observer, while negative $z$ values correspond to more distant stars. The systematic shift of the barycenters across slices reveals a coherent distance gradient across the LMC, consistent with an inclined disk geometry.}
    \label{fig:Slices}
\end{figure*}

\end{appendix}

\end{document}